\documentclass{article}
\usepackage{arxiv}

\usepackage[utf8]{inputenc} 
\usepackage[T1]{fontenc}    
\usepackage{hyperref}       
\usepackage{url}            
\usepackage{booktabs}       
\usepackage{amsfonts}       
\usepackage{nicefrac}       
\usepackage{microtype}      
\usepackage{array} 
\usepackage{graphicx}
\usepackage{natbib}
\usepackage{doi}
\usepackage{array}
\usepackage{amsmath}
\usepackage{float}
\title{From Search to Reasoning:\\ A Five-Level RAG Capability Framework for Enterprise Data}
\author{Corvic~AI Inc.}

\author{
    \bfseries Gurbinder Gill \quad Ritvik Gupta \quad Denis Lusson \quad Anand Chandrashekar \quad Donald Nguyen \\
    \textit{Corvic AI} \\
    \texttt{\{gill, ritvik, denis, anand, ddn\}@corvic.ai}
}

\renewcommand{\shorttitle}{From Search to Reasoning}

\renewcommand{\headeright}{Technical Report}

\renewcommand{\undertitle}{Technical Report}
\date{} 

\hypersetup{
pdftitle={From Search to Reasoning:\texorpdfstring{\\}{} A Five-Level RAG Framework for Enterprise Data},
pdfauthor={Corvic~AI},
}

\usepackage{tabularx}
\usepackage{multirow}
\usepackage{ragged2e}
\usepackage[table]{xcolor}

\newcolumntype{L}{>{\RaggedRight\arraybackslash}p{2.6cm}}
\newcolumntype{Y}{>{\RaggedRight\arraybackslash}X}

\definecolor{rowBlue}{HTML}{EAF4FF}
\definecolor{rowYellow}{HTML}{FFF7DB}
\definecolor{rowPink}{HTML}{FFECEC}
\definecolor{rowMint}{HTML}{E9F7EF}

\begin{document}
\maketitle

\begin{abstract}
Retrieval-Augmented Generation (RAG) has emerged as the standard paradigm for answering questions on enterprise data. Traditionally, RAG has centered on text-based semantic search and re-ranking. However, this approach falls short when dealing with questions beyond data summarization or non-text data. This has led to various attempts to supplement RAG to bridge the gap between RAG, the implementation paradigm, and the question answering problem that enterprise users expect it to solve.

Given that contemporary RAG is a collection of techniques rather than a defined implementation, discussion of
RAG and related question-answering systems benefits from a problem-oriented understanding.

We propose a new classification framework (L1--L5) to categorize systems based on data modalities and task complexity of the underlying question answering problems: L1 (Surface Knowledge of Unstructured Data) through L4 (Reflective and Reasoned Knowledge) and the aspirational L5 (General Intelligence). We also introduce benchmarks aligned with these levels and evaluate four state-of-the-art platforms: LangChain, Azure~AI Search, OpenAI, and Corvic~AI. Our experiments highlight the value of multi-space retrieval and dynamic orchestration for enabling L1--L4 capabilities. We empirically validate our findings using diverse datasets indicative of enterprise use cases.
\end{abstract}

\section{Introduction}
Large Language Models (LLMs) have introduced a new paradigm \citep{brown2020languagemodels} in data processing and utilization, and in enterprise environments, where private data represents a significant untapped asset, LLMs promise to transform vast stores of yet unexploited data into actionable insights. The potential applications are wide-ranging: from enhancing automated customer interactions to facilitating sophisticated decision-making processes. And, by leveraging a deeper understanding of data through LLMs, enterprises can augment their workforce with AI automation, thereby boosting productivity and operational efficiency.

However, the use of LLMs in enterprise scenarios has challenges. One significant hurdle is the tendency of LLMs to generate ``hallucinations'' or content-fabricated information not present in the input data \citep{huang2025surveyhallucinations}. This issue of hallucination, coupled with data incompleteness, can lead to unreliable outputs that jeopardize business decisions with potentially catastrophic consequences. As enterprises rely heavily on data integrity, the prevalence of misinformation generated by LLMs presents an existential risk.

To address these challenges, recent focus has shifted towards mitigating the hallucination phenomenon. Retrieval-Augmented Generation (RAG) \citep{lewis2020retrieval} has emerged as a promising approach, enhancing LLMs by integrating them with contextually relevant, enterprise-specific data \citep{izacard2021fid, karpukhin2020dense}. This integration aims to ground an LLM's responses with known facts and reduce the probability of fabrication \citep{shuster2021retrievalaugmentation}. Yet, despite its potential, RAG still grapples with the same challenges as using LLMs without RAG \citep{gao2024retrievalaugmentedgenerationlargelanguage}, such as maintaining response accuracy and relevance. This paper studies these issues, exploring innovative strategies to harness the full potential of LLMs in enterprise settings while ensuring accuracy and reliability crucial for business applications.

\begin{figure}[th]
\centering
\includegraphics[width=1\linewidth]{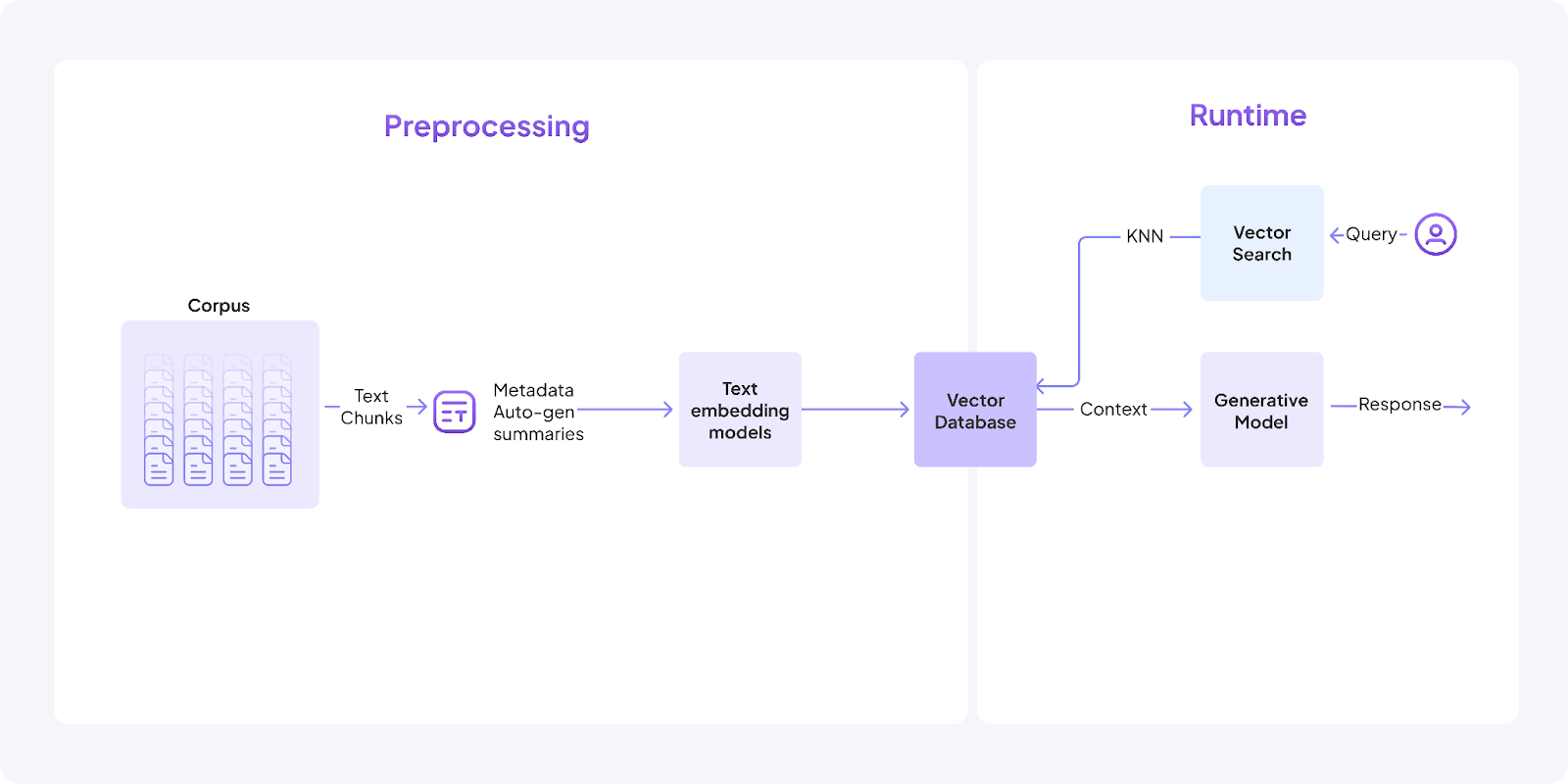}
\caption{\textbf{Generic RAG pipeline:} Architecture for L1 (surface knowledge of unstructured data) pipeline.}
\label{fig:generic_rag}
\end{figure}
 
\begin{figure}[th]
\centering
\includegraphics[width=1\linewidth]{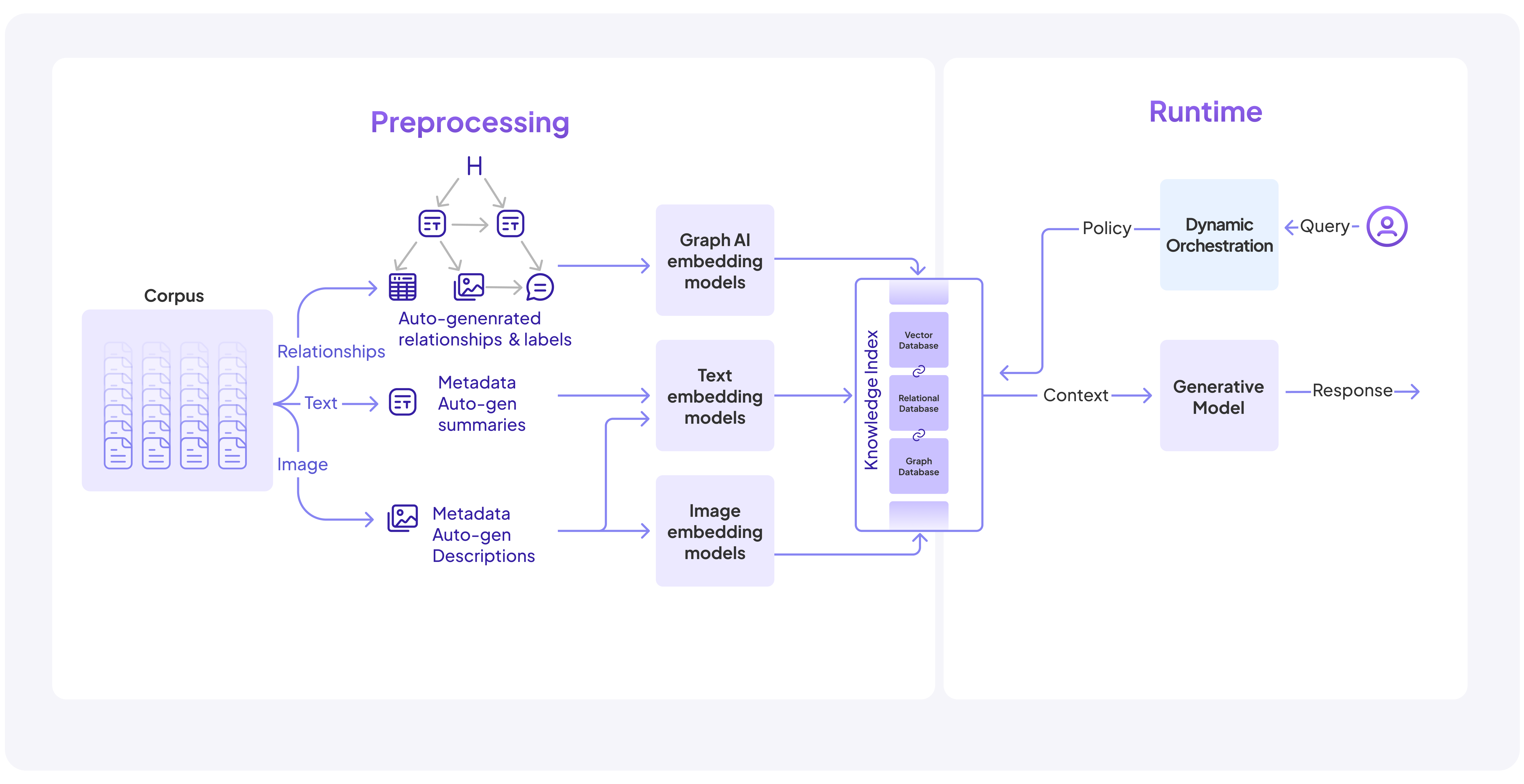}
\caption{\textbf{Corvic~AI:} Mixture of Spaces (MoS) and Adaptive Chain of Actions (ACoA) Architecture for L4 (reflective and reasoned knowledge) pipeline.}
\label{fig:corvic_rag}
\end{figure}

\section{Background}

Retrieval-Augmented Generation (RAG) (Fig.~\ref{fig:generic_rag}) has emerged as a
standard way to address the challenges of hallucination and incomplete responses of Large Language Models (LLMs) in data-sensitive environments. It adds a retrieval mechanism that fetches contextually relevant information from a private datasets. The original query is then augmented with this additional context before being passed to an LLM for question-answering~\citep{ram2023incontext}.
This method not only curtails the generation of hallucinated content but also enhances the completeness and factual accuracy of the responses~\citep{lewis2020retrieval} with respect to the private data. Despite its advancements, the RAG approach is still not a
panacea. For example, \cite{magesh2024hallucinationfree} have shown significant hallucination and incompleteness for legal questions, noting hallucination and incompleteness rates across systems between 10\% and 60\%. These findings have been observed across various other problem domains~\citep{amugongo2025raghealthcare}.

One limitation of RAG is that retrieval is through an embedding space that only captures language understanding and semantic similarity~\citep{xiong2021approximate}. While useful for
certain kinds of problems, this single-faceted approach
falls short when dealing with complex queries that require nuanced understanding and multi-faceted contextual grounding.
In an attempt to address this issue, the majority of the research in RAG systems has been focused on re-ranking algorithms~\citep{nogueira2020passagererankingbert, nogueira2019multistagedocumentrankingbert,zhuang2021tilde}, which refine the selection of retrieved documents based on their perceived relevance to the input query~\citep{guu2020realm, karpukhin2020dense}.

However, the efficacy of re-ranking is inherently limited by the quality of the initial retrieval phase. If the retrieval component fails to bring forward the relevant context, even the most sophisticated re-ranker cannot compensate for the initial deficiency. This decoupling of retrieval and re-ranking phases often leads to an accuracy bottleneck~\citep{yu2023generate}, which highlights the need for a different approach. Enhancing the retrieval mechanisms to consider multiple dimensions of semantic relevance and context variability~\citep{khattab2020colbert, santhanam2022colbertv2} paves the way for more robust and reliable outputs from LLMs, particularly in complex and dynamic information environments.
Ultimately, an LLM output is only as good as its retrieved inputs \citep{shi2023replug}. If pertinent information cannot be retrieved, no amount of LLM sophistication can compensate.

\section{Capability Levels: A Five-Level Framework}
While recent advancements in Retrieval-Augmented Generation (RAG) have delivered encouraging results, we argue that a more meaningful framing for enterprise use cases is: \textbf{What kinds of questions can be reliably answered from private data?} In enterprise settings, private data is not a free variable, it must be acquired, maintained, and protected. Much of the knowledge required for enterprise decision-making exists only in proprietary sources. Therefore, RAG capabilities should be measured not by end accuracy \textit{per se} but by the depth and reliability of answers across increasingly complex data landscapes.

To standardize this perspective, we propose a five-level classification system organized around increasing complexity of reasoning and understanding required to support real-world enterprise tasks.

\subsection{L1: \textit{Surface Knowledge of Unstructured Data}}
At L1, systems operate on unstructured text and perform basic retrieval using semantic similarity or keyword matching. They can answer factoid or lookup-style questions when the information exists explicitly in the source. However, they are unaware of any structural or contextual relationships among facts and cannot resolve ambiguity or synthesize across documents. These systems are effective for simple use cases such as semantic search, FAQ answering, and document summarization, where the required knowledge is shallow and localized.

\subsection{L2: \textit{Surface Knowledge of Multifaceted Data}}
L2 systems can process a wider range of multi-modal and multi-structural data types, including tabular data, knowledge graphs, text, images, and metadata. They incorporate awareness of document structure and modality, enabling better matching for more complex formats. However, their responses remain surface-level: they retrieve based on local matches without deeper inference or understanding. These systems are context-aware, but not reasoning-capable. They are suitable for moderately structured content like contracts, handbooks, or technical manuals, where surface context is helpful but not sufficient for deep understanding.

\subsection{L3: \textit{Implicative Knowledge of Multifaceted Data}}
L3 introduces reasoning capabilities. Systems at this level can synthesize information across multiple sources, infer relationships between disparate data points, and respond to queries that involve ambiguity, inconsistency, or incompleteness. They move beyond surface matching to link entities, draw conclusions across documents, and bridge questions with their many possible interpretations. This level is essential for enterprise scenarios involving diagnostic analysis, compliance validation, or decision support where answers depend on diverse, distributed, and contextualized knowledge.

\subsection{L4: \textit{Reflective and Reasoned Knowledge}}
At L4, the system becomes self-aware in its reasoning process. It can reflect on the sufficiency and reliability of its own answers, identify missing context, and revise its retrieval or reasoning strategy accordingly. It may dynamically invoke tools (e.g., calculators, database queries, visual parsers) or select from multiple data sources depending on the task. This level introduces true agentic behavior: the ability to plan, adapt, and reason recursively. It supports high-stakes applications such as enterprise copilots, compliance agents, and automated report generation where accuracy, traceability, and adaptive reasoning are non-negotiable.

\subsection{L5: \textit{General Intelligence}}
L5 represents an aspirational future state: a system capable of general-purpose reasoning across any domain, modality, or task type. It would possess the capacity to generalize knowledge, solve novel problems, and operate with minimal task-specific tuning or prompting. While it remains beyond current system capabilities, it serves as a direction for long-term research and system design.

\begin{table}[ht!]
\centering
\small
\begin{tabular}{
    p{0.3cm} 
    >{\raggedright\arraybackslash}p{2.5cm} 
    >{\raggedright\arraybackslash}p{3.5cm} 
    >{\raggedright\arraybackslash}p{4.5cm}
    >{\raggedright\arraybackslash}p{3.5cm}
    }
\toprule
\multicolumn{2}{l}{\textbf{Capability}}  & \textbf{Data Types Handled} & \textbf{Knowledge Capability Description} & \textbf{Example Tasks} \\
\midrule
L1 & Surface Knowledge of Unstructured Data & Primarily text-based unstructured sources & Retrieves explicit facts from text using keyword or semantic similarity; no structural or multi-source reasoning. & FAQ answering, document search, basic summarization \\
\midrule
L2 & Surface Knowledge of Multifaceted Data & Unstructured + semi-structured (tables, metadata, images) & Retrieves from diverse formats with structural context awareness, but answers remain surface-level without inference or synthesis. & Table lookup, contract clause search, metadata-based filtering \\
\midrule
L3 & Implicative Knowledge of Multifaceted Data & Unstructured + semi-structured + structured (databases, graphs) & Synthesizes across sources and modalities, infers relationships, and handles ambiguity, incompleteness, and inconsistencies. & Root cause analysis, compliance checks, product comparisons \\
\midrule
L4 & Reflective and Reasoned Knowledge & Multi-modal (text, tables, APIs, images, code) & Adapts retrieval strategies, orchestrates tools, and verifies or revises answers for high-stakes tasks: self-aware reasoning & Regulatory report generation, enterprise copilots, design validation \\
\midrule
L5 & General Intelligence & Any domain, modality, or data structure & Aspirational: domain-agnostic, autonomous reasoning and general-purpose problem-solving across all modalities. & Open-ended research, autonomous decision-making, novel problem solving \\
\bottomrule
\end{tabular}
\vspace{0.5em}
\caption{\textbf{Five-Level Knowledge Capability Framework for Enterprise RAG Systems.} Classification of systems based on the depth of reasoning and synthesis achievable from enterprise data, along with representative tasks at each level.}
\label{tab:rag_knowledge_levels}
\end{table}

\section{Methodology: Enabling L4 Systems}
Achieving \textbf{L4 (reflective and reasoned knowledge)} requires a fundamental departure from traditional dense-vector search and static chunk-based retrieval. 
While L1 through L3 are sufficient for simple lookup and context-aware or even cross-source synthesis, enterprise environments often
demand capabilities that are dynamic, multi-modal, and self-correcting. L4 systems must retrieve relevant information but also evaluate, adapt, and orchestrate their reasoning to ensure accuracy and completeness in mission-critical settings.

To meet these demands, an L4 system must:
\begin{itemize}
    \item Accurately represent and unify complex \textbf{structured}, \textbf{semi-structured}, and \textbf{unstructured} data
    \item Retrieve context across multiple semantic, structural, and metadata dimensions
    \item Dynamically adapt retrieval and reasoning strategies to the query's intent
    \item Seamlessly orchestrate multiple tools, models, and knowledge views
    \item Self-reflect, re-plan, and re-execute retrieval steps to ``connect the dots''
\end{itemize}

We introduce \textbf{Corvic~AI}, a platform designed from the ground up to operate at L4. Corvic~AI integrates three pillars, (1) \textbf{Structure-Aware Data Representation}, (2) \textbf{Mixture of Spaces}, and (3) \textbf{Adaptive Chain of Actions}, transforming static pipelines into highly accurate agentic reasoning systems.

\subsection{Structure-Aware Data Representation}
Enterprise knowledge ecosystems rarely exist in a single format. They blend unstructured content (e.g., manuals, reports, narrative text) with structured elements (e.g., tables, forms, relational databases, graphs). L4 performance depends on a unified representation that captures the multi-faceted nature of the data.

The \textbf{Structure-Aware Data Representation} approach parses each document into an enriched intermediate form encoding:
\begin{itemize}
    \item Document hierarchies, e.g., section, subsection, heading, paragraph hierarchies
    \item Embedded tables, lists, and field-level forms
    \item Local and global metadata, e.g., document type, version, author
    \item Cross-references, e.g., ``See Table 1'', ``as shown above''
\end{itemize}

This structure-aware representation bridges the gap between unstructured and structured modalities, enabling precise section or field-specific retrieval, cross-format linking, and context mapping across disparate data types. It is modular and extensible, and it forms the foundation for the techniques that follow (as shown in Fig.~\ref{fig:corvic_rag}).

\subsection{Mixture of Spaces}
Traditional RAG systems (as shown in Fig.~\ref{fig:generic_rag}) compress all content into a single semantic vector space, losing critical structural and contextual cues. The \textbf{Mixture of Spaces} (MoS) approach builds multiple parallel representations of the same document:
\begin{itemize}
    \item A \textit{semantic space} for meaning-rich embeddings
    \item A \textit{structural space} modeling layout, hierarchy, and relationships
    \item A \textit{metadata space} encoding titles, tags, and annotations
\end{itemize}

This multi-view indexing allows the system to retrieve relevant context through different pathways. For example, if a semantic search misses a passage, a structural or metadata search may still retrieve it based on its heading or relational position. This redundancy increases both recall and precision. This is key for high-stakes decision-making.

\subsection{Adaptive Chain of Actions}
Most RAG pipelines follow a rigid retrieve, augment, and generate sequence. In contrast, the \textbf{Adaptive Chain of Actions} (ACoA) framework dynamically assembles query-specific retrieval and reasoning plans.

Adaptive Chain of Actions can:
\begin{itemize}
    \item Select the optimal retrieval spaces based on query intent
    \item Sequence multiple retrieval and enrichment steps (e.g., schema lookup $\rightarrow$ graph traversal $\rightarrow$ synthesis)
    \item Integrate specialized tools for computation, visualization, or external data access
\end{itemize}

By adapting its strategy on the fly, ACoA supports reflective reasoning: the ability to detect gaps, revise plans based on discovered knowledge, and re-execute and refine steps until the answer meets completeness and reliability criteria.

In combination, Structure-Aware Data Representation, Mixture of Spaces, and Adaptive Chain of Actions enable Corvic~AI to operate at \textbf{L4 (reflective and reasoned knowledge)}, bridging structured and unstructured sources, adapting to varied enterprise queries, and delivering accurate, explainable, and high-trust outputs (as shown in Fig.~\ref{fig:corvic_rag}). The following section evaluates Corvic~AI's performance against other leading RAG platforms.

\section{Experimental Results}

\subsection{Datasets}
Conventional benchmarks for Retrieval-Augmented Generation (RAG) typically evaluate performance on preprocessed, clean text chunks. This approach, however, is not representative of real-world enterprise scenarios, where information is predominantly stored in unstructured documents. For a more precise performance assessment on realistic enterprise tasks, our evaluation employs four datasets with source documents in PDF format. Each dataset is paired with a corresponding question corpus, formulated to probe for information embedded within the documents' complex structure~\footnote{All datasets are hosted on HuggingFace for convenient download and use (see Table~\ref{tab:dataset_sizes}).}.

\begin{itemize}
    \item \textbf{DelucionQA:} The \texttt{DelucionQA}\footnote{\url{https://github.com/boschresearch/DelucionQA}} dataset \citep{sadat2023delucionqa} is a benchmark for question answering, using the Jeep 2023 Gladiator Car manual as its knowledge base. DelucionQA provides a challenging dataset designed to evaluate a system's ability to answer specific, technical questions based on a complex, real-world document.

    \item \textbf{FinTabNet:} The \texttt{FinTabNet} dataset is a large-scale collection of PDFs from the financial earnings reports of Fortune 500 companies~\citep{zhao2024tabpedia, zheng2020fintabnet}. Released as an open-source resource by IBM Research, it is freely available on the IBM Developer Data Asset Exchange.
    
    \item \textbf{DMV Handbooks:}  The DMV Handbooks dataset is a Corvic~AI-created dataset comprising the official DMV handbooks from all 50 US states. This collection of 50 long-form, structurally complex PDFs serves as the source for answering comprehensive sets of official state DMV test questions. It evaluates a system's ability to perform precise information retrieval across a large and heterogeneous document collection, mirroring a common enterprise requirement.
    
    \item \textbf{Architectural Manuals:} The Architectural Manuals dataset is a Corvic~AI-created dataset containing real-world complex architectural and engineering documentation. This includes CAD drawings, technical schematics, circuit diagrams, and building project blueprints that combine intricate 2D visual elements with embedded technical text. The dataset is designed to test a system’s ability to jointly process and reason over multi-modal inputs where precise visual–text alignment is critical.

\end{itemize}

\begin{table}[ht!]
\centering
\small
\begin{tabular}{l p{2.4cm} c c c p{4.5cm}}
\toprule
\textbf{Dataset} & \textbf{Modality} & \textbf{\#Q} & \textbf{\#Files} & \textbf{\#Pages} & \textbf{Enterprise Use Case} \\
\midrule
DelucionQA & Text & 184 & 1 & 384 & Basic retrieval from manuals/product guides; customer support and FAQ automation. \\
\midrule
FinTabNet & Text + Tables & 1,000 & 591 & 591 & Financial analysis from reports; risk assessment, KPI computation, and compliance reporting. \\
\midrule
DMV Handbooks & Text + Images + Tables & 135 & 50 & 4,689 & Regulatory/compliance document search; onboarding, training, and policy lookup. \\
\midrule
Architecture & Images + Diagrams + Embedded Text & 43 & 8 & 36 & Engineering/design review; reasoning over blueprints, CAD diagrams, and technical schematics. \\
\bottomrule
\end{tabular}
\vspace{0.5em}
\caption{\textbf{Datasets (questions, number of PDF files and pages) in increasing complexity.} Complexity increases with modality variety, reasoning depth, and retrieval difficulty. The datasets are available for download at: \url{https://huggingface.co/datasets/corvicai/delucionqa}, 
\url{https://huggingface.co/datasets/corvicai/FinTabNet_ComTQA},
\url{https://huggingface.co/datasets/corvicai/dmv-handbooks}, and \url{https://huggingface.co/datasets/corvicai/architectural-manuals}.}
\label{tab:dataset_sizes}
\end{table}

\begin{table}[h!]
\centering
\small
\setlength{\tabcolsep}{4pt}
\renewcommand{\arraystretch}{1.18}
\begin{tabularx}{\textwidth}{@{} p{0.9cm} L Y Y Y @{}}
\toprule
 & \textbf{Dataset} & \textbf{Example Query} & \textbf{Ground Truth} & \textbf{Complexity} \\
\midrule
\rowcolor{rowBlue}
L1 & \multirow{1}{*}{\textsc{DelucionQA}} 
  & What is the DEF?
  & \texttt{Diesel Exhaust Fluid}
  & Simple question on simple text search; answer explicitly stated in plain text. \\
\midrule
\rowcolor{rowYellow}
L2 & \textsc{FinTabNet} 
  & For PPL, which portfolio had the highest percentage of assets allocated to debt securities in 2015?
  & \texttt{Growth Portfolio} (\textit{Rationale: Debt Securities = 13\%}) 
  & Direct question but requires searching and reading tables from unstructured data accurately to locate the answer. \\
\midrule
\rowcolor{rowPink}
L3 & \textsc{DMV Handbooks} 
  & What is the estimated BAC for a 120-pound woman in California after 2 drinks?
  & \texttt{$\approx$0.11\%} (\textit{Rationale: BAC chart entry with 120 lb, 2 drinks})
  & Direct question but answer is not in plain text; requires combining table lookup with multimodal (chart) parsing and implicit knowledge generation. \\
\midrule
\rowcolor{rowMint}
L4 & \textsc{Architectural Manuals} 
  & Find an enclosure with three knockouts.
  & \texttt{Enclosed Series KT7 Motor Controller} (\textit{Rationale: From Figure~3, the enclosure with three circular knockouts}) 
  & Indirect question requiring deeper domain understanding; needs multimodal retrieval and reasoning across diagrams, labels, and text. \\
\bottomrule
\end{tabularx}
\vspace{0.35em}
\caption{\textbf{Illustrative examples mapped to capability levels.} The additional complexity column explains the reasoning complexity behind each example.}
\label{tab:rag_examples_levels_reasoning}
\end{table}

The examples in Table~\ref{tab:rag_examples_levels_reasoning} illustrate the progressive complexity of retrieval-augmented generation (RAG) tasks as the capability level increases. At L1, such as the DelucionQA dataset, the question is straightforward and the answer exists verbatim in the text, requiring only basic semantic search over unstructured content.  FinTabNet (L2) introduces structured elements, in this case, a table, which demand accurate parsing and retrieval of non-textual formats, but answers are still directly factual. DMV Handbooks (L3) raise the bar by requiring multi-modal comprehension: the system must interpret a chart, map its contents to the query, and apply implicit reasoning to extract the correct result. Finally, Architectural Manuals (L4) represent an indirect, domain-specific query where the answer is not explicitly stated in any single source; instead, it requires understanding technical terminology (``knockouts''), combining information from diagrams and text, and synthesizing a correct and contextually appropriate answer. This progression underscores how higher levels demand richer data representations, more sophisticated retrieval strategies, and reasoning capabilities that extend well beyond surface-level keyword or vector search.

\subsection{Evaluation}
For our experiments, we selected four representative implementations aligned with the capability levels. \textbf{LangChain} was used to implement an {L1 (surface knowledge of unstructured data)} system, focused on semantic search over unstructured text. \textbf{OpenAI RAG} and \textbf{Azure AI Search} represent {L2 (surface knowledge of multifaceted data)} systems, capable of handling unstructured and semi-structured data with surface-level structural awareness but without deep reasoning. \textbf{Corvic~AI} is representative of an {L4 (reflective and reasoned knowledge)} system. It incorporates structure-aware representation, Mixture of Spaces, and Adaptive Chain of Actions for reflective and adaptive reasoning.

To ensure consistent comparison of response quality, we used \textbf{gpt-4.1} for the final answer generation and summarization step. Depending on the system, there are a variety of parameters that affect retrieval. To obtain representative
behavior, we used the default retrieval settings.
LangChain and Corvic~AI use OpenAI's \textbf{text-embedding-3-large} embeddings, while OpenAI RAG uses its native embedding solution. LangChain uses \textbf{ChromaDB} as the vector database. For Azure AI Search, we used the default parameters which use OpenAI's \textbf{text-embedding-3-small} model for embeddings.

For evaluation, we employed an LLM-as-a-judge \citep{zheng2023llmjudge} approach using the \texttt{ragas}\footnote{\url{https://github.com/explodinggradients/ragas}} library \citep{es024ragas}, leveraging \textbf{gemini-2.0-flash} as the evaluation model to score the outputs. This choice ensures an unbiased assessment, as the judge model is from a different family than the generation model used in the pipelines. 

\textbf{Note:} In this study, our primary goal was to measure the performance advantage that L4 capabilities bring to a knowledge search and management use case primarily involving unstructured data; future work will extend this evaluation to multi-structured datasets along with unstructured ones.

\begin{table}[th!]
\centering
\begin{tabular}{l c c c c}
\toprule
\textbf{Framework} & \textbf{DelucionQA} & \textbf{FinTabNet} & \textbf{DMV Handbooks} & \textbf{Architectural Manuals} \\
\midrule
LangChain & 69.83 & 38.23 & 64.63 & 36.63 \\
\midrule
Azure AI Search & 67.11 & 41.75 & 51.30 & 40.12 \\
\midrule
OpenAI RAG & 67.53 & 52.55 & 71.85 & 58.72 \\
\midrule
Corvic AI & \textbf{82.74} & \textbf{63.83} & \textbf{79.44} & \textbf{65.12} \\
\bottomrule
\end{tabular}
\vspace{0.35em}
\caption{A performance comparison of four AI systems across different datasets. Scores are reported as accuracy percentages (\%) with the highest score in each row shown in bold.}
\label{tab:evaluations}
\end{table}

\begin{figure}[ht!]
    \centering

    \begin{minipage}{0.45\textwidth}
        \centering
        \includegraphics[width=\linewidth]{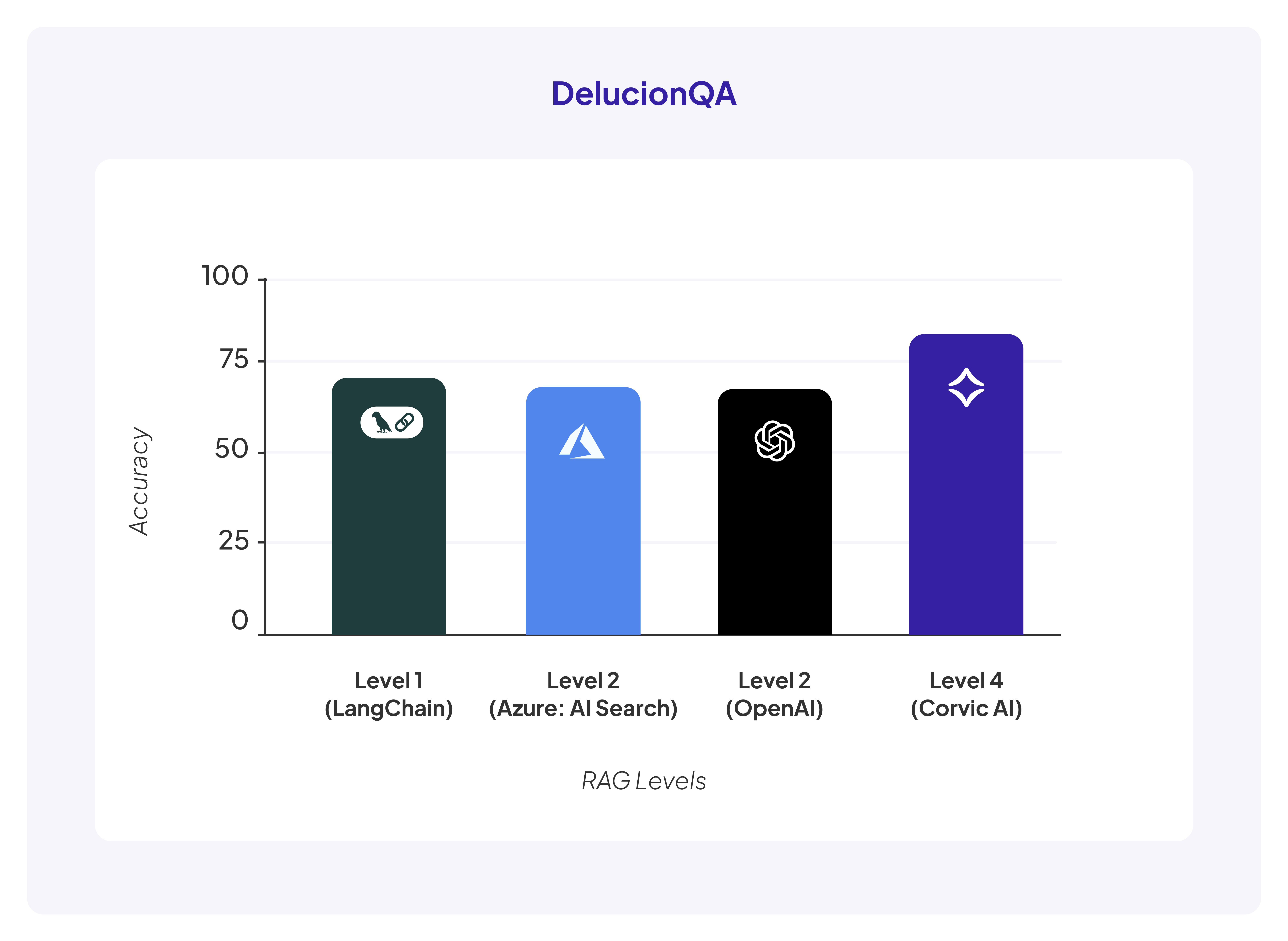}
        \caption{\textbf{DelucionQA:} Manual-like Q\&A over unstructured text and images. Corvic~AI leads across basic retrieval questions on a single long manual.}
    \end{minipage}
    \hfill
    \begin{minipage}{0.45\textwidth}
        \centering
        \includegraphics[width=\linewidth]{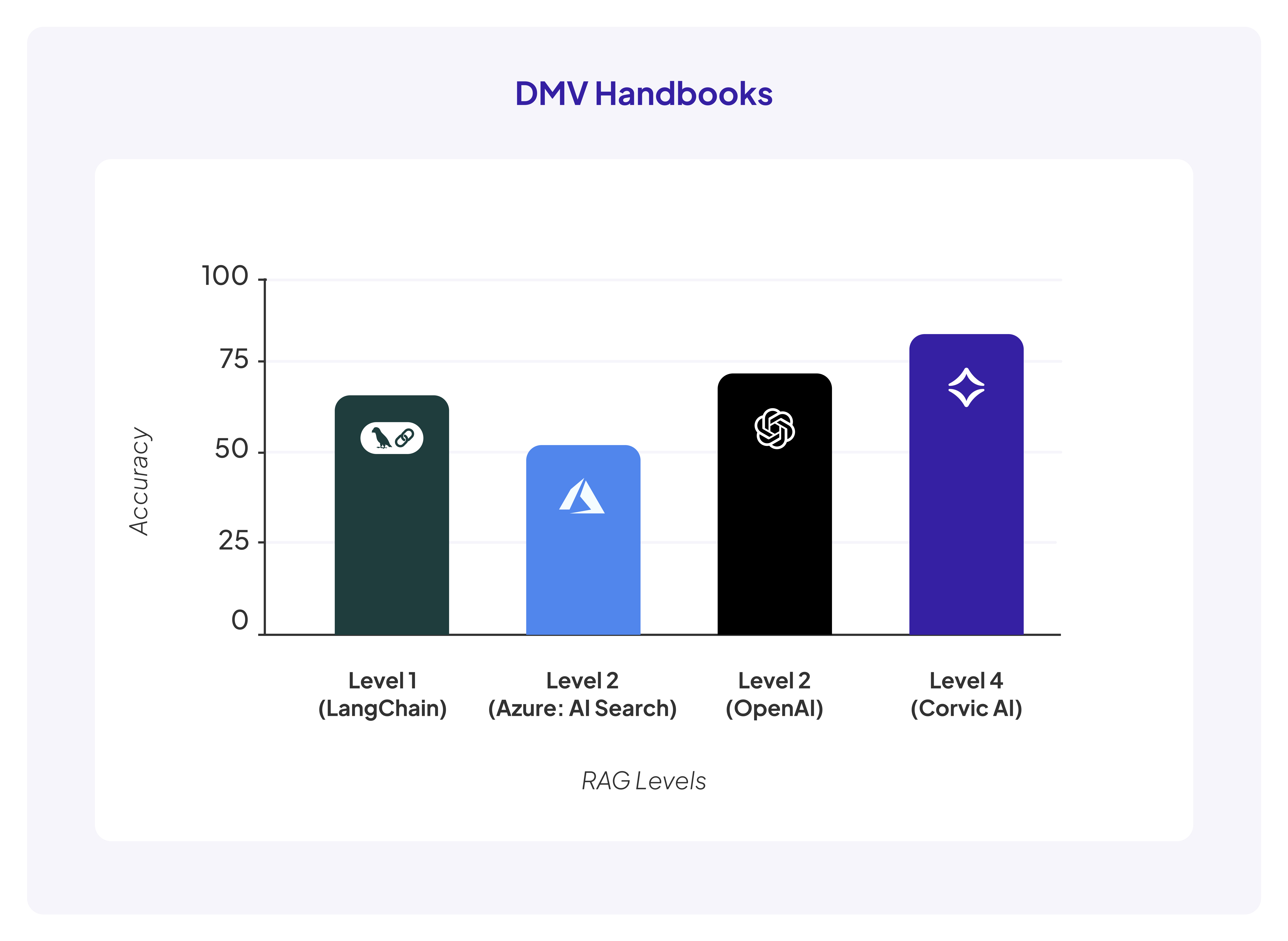}
        \caption{\textbf{DMV Handbooks:} Mixed text, tables, and images. Corvic~AI widens the gap as queries require cross-section references and light reasoning.}
    \end{minipage}

    \vspace{0.5cm}

    \begin{minipage}{0.45\textwidth}
        \centering
        \includegraphics[width=\linewidth]{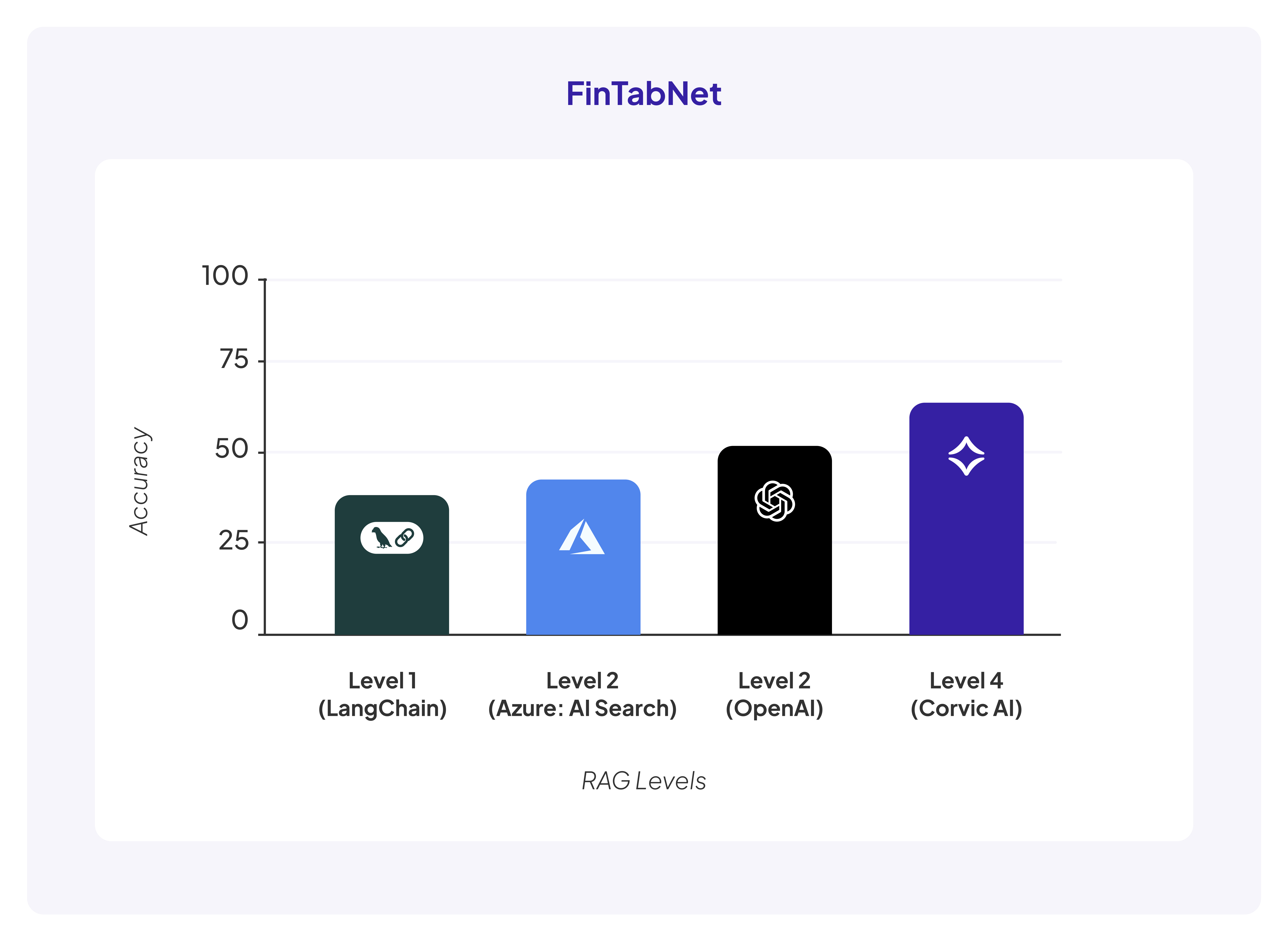}
        \caption{\textbf{FinTabNet:} Table-centric questions. Corvic~AI outperforms on multi-step, numeric, and schema-aware reasoning over financial tables.}
    \end{minipage}
    \hfill
    \begin{minipage}{0.45\textwidth}
        \centering
        \includegraphics[width=\linewidth]{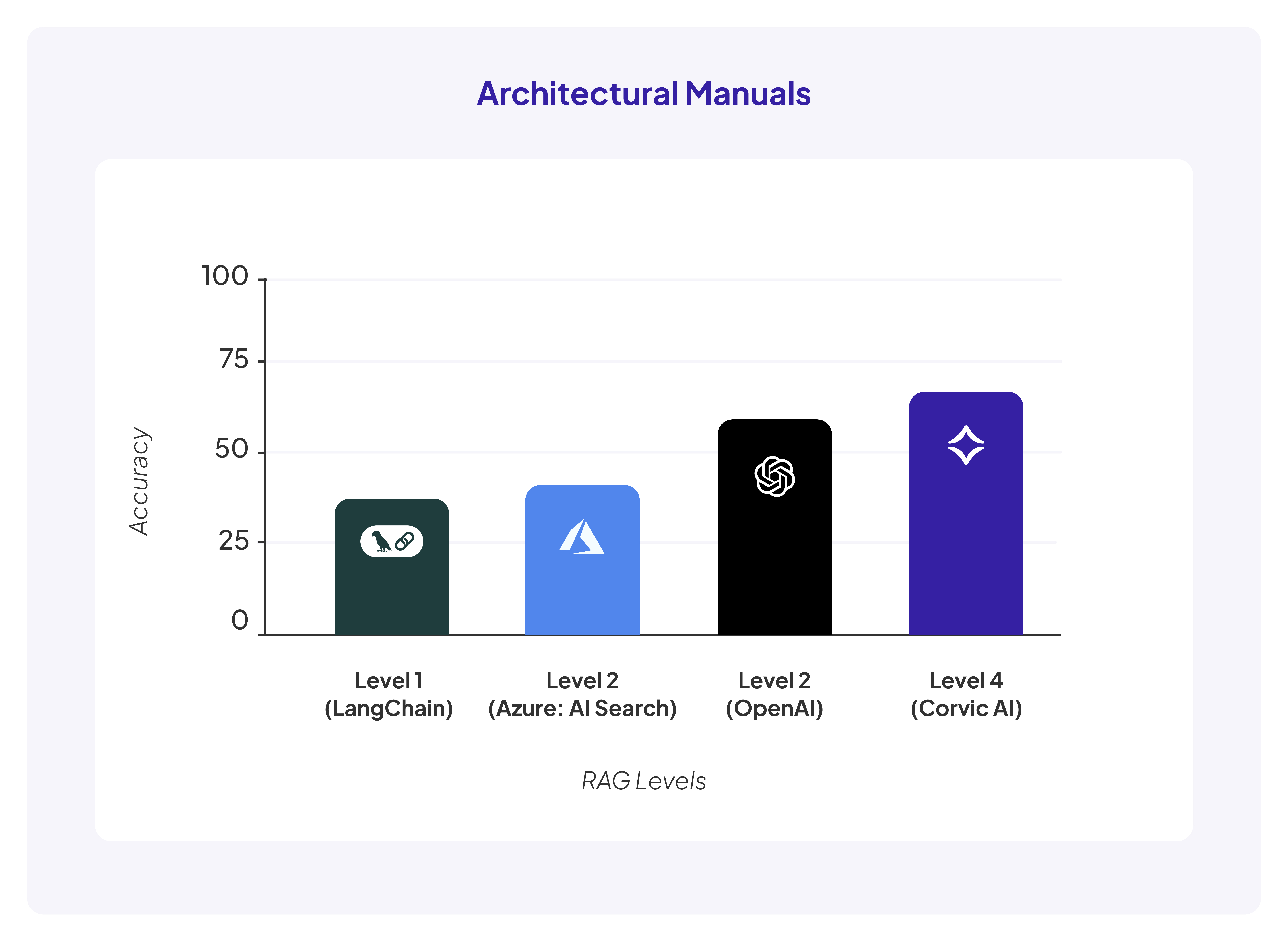}
        \caption{\textbf{Architectural Manuals:} Diagrams with embedded text and layout noise. Corvic~AI shows gains on visual–text fusion and context-heavy queries.}
    \end{minipage}

\caption{\textbf{Performance of systems aligned to capability levels across four datasets.}
Bars show comparative scores per system for each dataset (higher is better).
Across all corpora, the \textbf{L4 (reflective and reasoned knowledge)} system, Corvic~AI leads; 
\textbf{L2} varies by implementation (e.g., retrieval strategy, embeddings, re-rankers) and gaps widen with dataset complexity (DelucionQA $<$ DMV $<$ FinTabNet $<$ Architecture).}
    \label{fig:all_plots}
\end{figure}

\subsection{Key Observations}
Across all four datasets, Corvic~AI, which is designed to solve \textbf{L4 (reflective and reasoned knowledge)} problems, delivers the strongest accuracy, even when the corpora are primarily unstructured. The combination of Structure-Aware Data Representation, Mixture of Spaces, and Adaptive Chain of Actions consistently recovers evidence that single-space pipelines miss (e.g., passages discoverable via headings or layout structure rather than pure semantics) and then verifies or augments that evidence through iterative retrieval. In practice, this manifests as higher groundedness and fewer omissions on long, layout-heavy PDFs, where relevant content is often contextually implied by section hierarchy or table locality rather than surface phrasing.

Within the same capability tier, performance can vary markedly, e.g., among the two \textbf{L2 (surface knowledge of multifaceted data)} systems (Azure AI Search vs.\ OpenAI RAG). This divergence highlights the impact of design choices: vector vs.\ keyword vs.\ hybrid retrieval (and their weighting), domain vs.\ general-purpose embedding models, cross-encoder vs.\ bi-encoder re-rankers, and the summarization LLM's length control and citation style. Small configuration shifts can change which passages are retrieved, how they are ordered, and what ultimately makes it into the context window. This, in turn, leads to materially different answer quality without changing the nominal ``level'' of the system.

Dataset complexity acts as an amplifier. On simpler, manual-like material (DelucionQA), the gap between \textbf{L1 (surface knowledge of unstructured data)} and higher level systems is narrower because many answers are explicitly stated and semantically proximate. As structure and reasoning demands grow, from the simpler DMV (mixed modalities and long-range references) dataset to the increasingly more complicated FinTabNet (table-centric, numeric, and multi-step computation) and Architectural Manuals (visual–text fusion with layout noise) datasets, the advantage of Corvic~AI widens compared to L2 systems on average. Multi-space retrieval and reflective planning help recover table cells, respect schema and column context, and follow cross-figure references, while systems designed for lower levels suffer recall drops or over-summarization errors.

A related pattern is \textit{robustness}. Corvic~AI not only achieves higher accuracy but also exhibits smaller swings across datasets. Reflective re-planning reduces brittleness to corpus idiosyncrasies (e.g., varied headings, inconsistent table markup, OCR artifacts) by retrying with alternate spaces or tools when evidence is sparse. In contrast, systems designed for L1 and L2 show larger sensitivity to modality mix and layout density because they rely more heavily on a single retrieval view and fixed ordering.

Finally, the \textit{error taxonomy} differs by level. LangChain (L1) tends to miss cross-section or cross-document cues and may retrieve semantically similar but contextually wrong passages. L2 systems improve recall but are brittle: changes in hybrid weights or re-ranker choice can flip top passages, and table/figure evidence is often underutilized. Corvic~AI mitigates both classes of errors by (1) retrieving through multiple views (semantic, structural, metadata), (2) verifying coverage against the question intent, and (3) iterating when gaps are detected, which collectively reduces hallucinations and incomplete answers on enterprise-style documents.

\section{Discussion}
Our results show that techniques for addressing \textbf{L4 (reflective and reasoned knowledge)} problems deliver a clear and consistent advantage for enterprise knowledge search and management even when corpora are predominantly unstructured. The architectural ingredients behind L4 (Structure-Aware Data Representation, Mixture of Spaces, and Adaptive Chain of Actions) translate into higher accuracy and more reliable grounding across diverse datasets, while lower levels exhibit wider swings.

The study also clarifies how to interpret ``levels'' versus ``implementations''. A level specifies a \emph{capability envelope}. Where a concrete system lands within that envelope depends on design choices, e.g., retrieval strategy, embeddings, re-rankers, summarization LLM. Thus, the classification provides the ceiling and direction of travel, and implementation determines realized performance within a level.

Finally, as task and data complexity rise, the \emph{value} of higher levels grows. This suggests a pragmatic roadmap: organizations starting with simpler use cases can begin at lower levels but should plan upgrades toward L4 as modality mix, reasoning depth, and reliability requirements increase.

\textbf{\textit{Scope.}} This study focused on question answering on datasets primarily comprised of unstructured-data. Future work will extend to structured and multi-modal corpora and include L3 baselines and targeted ablations to map accuracy–latency–cost trade-offs more completely.


\section{Conclusion}
We introduced a five-level RAG classification that organizes system capabilities from \textbf{L1 (surface knowledge of unstructured data)} through to \textbf{L4 (reflective and reasoned knowledge)} and the aspirational \textbf{L5 (general intelligence)}. This is intended both as a methodological framework and as a practical guide for aligning enterprise requirements with implementation choices.

At lower levels, results reveal that choices of retrieval strategy, embeddings, re-rankers, and summarization LLMs can shift outcomes. This underscores the value of system classification for setting appropriate expectations.

\textbf{Corvic~AI} consistently outperforms representative L1 and L2 systems on knowledge search and management tasks, even on predominantly unstructured datasets. Since Corvic~AI was
designed to solve L4 problems, this highlights the leverage of
addressing complex problems upfront.

As enterprises adopt RAG for mission-critical workflows, framing the problem correctly becomes essential for determining the right system capabilities---and Corvic~AI exemplifies how higher level problem solving can deliver measurable enterprise value today.

\newpage
\bibliography{references}
\end{document}